\documentclass[twocolumn,a4paper]{article}
\usepackage{verbatim}
\usepackage{amssymb}
\usepackage{amsmath}
\usepackage{amsthm}
\usepackage{graphicx}
\usepackage{hyperref}
\usepackage{authblk}

\usepackage{etoolbox}
\AtBeginEnvironment{align}{\setcounter{subeqn}{0}}
\newcounter{subeqn} \renewcommand{\thesubeqn}{\theequation\alph{subeqn}}%
\newcommand{\subeqn}{%
  \refstepcounter{subeqn}
  \tag{\thesubeqn}
}

\title{Conservation of Optical Chirality in Superconductors as a measure for $5$ dimensional Electromagnetism}
\author[1]{H. Lashkari-Ghouchani\thanks{hadilashkari@gmail.com}}
\author[2]{M.H. Alizadeh\thanks{halizade@bu.edu}}
\affil[1]{\textit{Faculty of Science, PNUM. Mashhad Iran.}}
\affil[2]{\textit{Physics Department, Boston University, Boston, MA 02215}}
\affil[2]{\textit{Photonic Center, Boston University, Boston, MA 02215}}

\begin{document}
\maketitle
\makeatletter
\def\blfootnote{\xdef\@thefnmark{}\@footnotetext}
\makeatother
\blfootnote{Corresponiding author: hadilashkari@gmail.com}

\begin{abstract}
Currently only three spatial and one temporal dimensions are considered to be ``physical''. Recently, solutions to a plethora of questions have used the notion of extra-dimensions. The experimental verification of the existence of such extra dimensions, however, remains elusive. Here by applying standard electromagnetic theory to five dimensional Minkowski space-time, we propose a novel test ground to search for a new non-compactified lengthlike extra dimension. In this regard, we propose superconductors as a test bed for this hypothetical new extra dimension.
\end{abstract}

\section*{INTRODUCTION}
Mankind was always familiar with three spatial dimensions and time had to wait a long time to be introduced as a new Physical dimension by Einstein in his theory of \textit{Special Relativity}. Currently these dimensions are the only ones considered ``physical'' in its strict sense. Recently, however, a surging interest has revived in introducing new extra dimensions to solve a wide variety of problems.  Some most notable of these works try to tackle the hierarchy problem of \textit{Standard Model} \cite{csaki1999cosmology,randall1999large,cline1999cosmological,arkani1998hierarchy,agashe2005flavor}. Majority of such theories have Kaluza-Klein extra dimension, which was an attempt for unification of electromagnetism with \textit{General Relativity}\cite{kumar2006gravitons,bhattacherjee2011search,cornell2011evolution,datta2013vacuum}. Extensive reviews about different types of extra dimensions can be found here \cite{overduin1997kaluza,shiu1998tev,nihei1999inflation,kanti2004black}.  \\
Here we investigate a potential candidate to experimentally verify the existence of a new non-compactified lengthlike dimension in the context of electromagnetism. Nominally we focus on theories with five dimensions, one temporal, three
spatial and one hypothetical extra lengthlike dimension, HELD. We specifically propose new physical niches to look for such an extra dimension at. Similar approaches have been used in \cite{visser1999exotic,chodos1980has} which utilized non-compactified lengthlike extra dimensions. Unlike refs \cite{visser1999exotic} and \cite{chodos1980has} where particles were trapped on the fifth dimension, here we assume that they are falling on the extra dimension. \\
In the first section we derive $4$-dimensional Maxwell's equations as an introduction to our approach to derive similar equations for $5$D in section 2. Additionally, we explain why such an extra dimension is yet to be detected. In section 3 we will argue that type I superconductors may be the desirable niche for experimental verification of such an extra dimension. 

\section*{$4$D ELECTROMAGNETIC}
As a pedagogical prelude to our derivation of $5$-EM, in this section we review the equations and relations of standard electromagnetism. The \textit{Maxwell's equations} in $n$-\textit{Minkowski} space-time with $x^\mu$ coordinates are:
\begin{flalign}
\partial_{[\gamma} F_{\alpha\beta]}&=0\label{maxwellF}\\
\partial_\beta F^{\alpha\beta}&=4\pi J^\alpha\label{maxwellFJ}
\end{flalign}
where $F$ is the eletromagnetic tensor field and $J$ is the current-density vector. The metric of $n$-Minkowski space-time is declared by:
\begin{flalign}\label{minkmetric}
(\eta_{\alpha\beta})=\text{diag}(-v^2,1,1,\ldots,1)
\end{flalign}
where in this paper $t=x^0$ is time and $x,y,z=x^1,x^2,x^3$ are spatial coordinates. As long as the determinant of the metric is non-zero, by applying appropriate coordinates any constant metrics, $g_{\alpha\beta}$, can be reduced to \textit{Minkowski} metric, \eqref{minkmetric}, so we will generally use $g_{\alpha\beta}$ instead of \eqref{minkmetric}. This metric, \eqref{minkmetric}, allows the time measurement unit to be different from that of space. Since speed of light, $v$, is considered to be a constant, we restrict ourselves to monochromatic fields while dealing with dispersive media. \\
In this section, we look for $4$-EM fields. The lower indices, $F_{\alpha\beta}$, relate to $\mathbf{B}$ and $\mathbf{E}$, \eqref{maxwellF}, while the upper indices, $F^{\alpha\beta}$, relate to $\mathbf{H}$ and $\mathbf{D}$, \eqref{maxwellFJ}. In order to obtain the correct equations in $4$-Minkowski space-time, we choose the components of $F^{\alpha\beta}$ and $F_{\alpha\beta}$ as follows:
\begin{align}\label{FField1}
\big(F^{\alpha\beta}\big):=
\begin{pmatrix}
 0   &  D_x &  D_y &  D_z \\
-D_x &  0   &  H_z &- H_y \\
-D_y &- H_z &  0   &  H_x \\
-D_z &  H_y &- H_x &  0
\end{pmatrix}
\end{align}
and
\begin{align}\label{FField2}
\big(F_{\alpha\beta}\big):=\mu
\begin{pmatrix}
 0   & -E_x & -E_y & -E_z \\
 E_x &  0   &  B_z &- B_y \\
 E_y &- B_z &  0   &  B_x \\
 E_z &  B_y &- B_x &  0
\end{pmatrix}
\end{align}
where $\mathbf{B}$, $\mathbf{H}$, $\mathbf{E}$ and $\mathbf{D}$ are defined by the components of these matrices. $\mu$ is a constant. By this definitions, the relations of $\mathbf{B}$ and $\mathbf{H}$ or $\mathbf{E}$ and $\mathbf{D}$ become invariant for any metric, 
\begin{align}\label{FRelations}
F^{\alpha\beta}=g^{\alpha\mu}g^{\beta\nu}F_{\mu\nu}.
\end{align} 
Here, we choose a metric, which is \textit{conformal} to the $4$-Minkowski's metric, \eqref{minkmetric}, namely
\begin{flalign}\label{metric}
(g_{\alpha\beta}):=\mu(\eta_{\alpha\beta})=\text{diag}(-\frac{1}{\epsilon},\mu,\mu,\mu)
\end{flalign}
where $v=1/\sqrt{\mu\epsilon}$. As a result \eqref{FRelations} reduced to $\mathbf{B}=\mu \mathbf{H}$ and $\mathbf{E}=\mathbf{D}/\epsilon$ which are the electromagnetic constitutive equations. $\mu$ is the magnetic permeability and $\epsilon$ is the electric permittivity. So by using \eqref{maxwellF} and \eqref{maxwellFJ}, the correct $4$-Maxwell's equations take the form
\begin{flalign}\label{maxwell1}
\nabla.\mathbf{B}&=0\nonumber\\
\partial_t\mathbf{B}+\nabla\times \mathbf{E}&=0\nonumber\\
\end{flalign}
and
\begin{flalign}\label{maxwell2}
\nabla.\mathbf{D}&=4\pi J^0\nonumber\\
-\partial_t\mathbf{D}+\nabla\times \mathbf{H}&=4\pi\mathbf{J}.\nonumber\\
\end{flalign}
These are ordinary definitions of Maxwell's equations. In the next section we will use this method to find generalization of Maxwell's equations to $5$-EM from \eqref{maxwellF} and \eqref{maxwellFJ}. It is worth mentioning that definitions, \eqref{FField1} and \eqref{FField2}, are consistent with the motion of a charged particle in electric and magnetic fields. Let $p^\alpha=mdx^\alpha/d\tau$ be the momntum of the particle where $m$ is its mass and $\tau$ is its locale time then
\begin{flalign}\label{lorantz}
\frac{dp^\alpha}{d\tau}=q{F^\alpha}_\beta u^\beta
\end{flalign}
are the equations of motion of the charged particle, or \textit{Lorentz law}, where $q$ is particle's charge. By expanding \eqref{lorantz}, using \eqref{FField1}-\eqref{metric}, we can find its original format, which shows the consistency.\\
Another important equation that we will generalize in the next section is the continuity equation for \textit{optical chirality}.
\begin{flalign}
&\partial_t\big(\mathbf{B}.\partial_t\mathbf{D}-\mathbf{D}.\partial_t\mathbf{B}\big)
+\nabla.\big(\mathbf{E}\times\partial_t\mathbf{D}+\mathbf{H}\times\partial_t\mathbf{B}\big)\nonumber\\
&=4\pi (\mathbf{J}.\partial_t\mathbf{B}-\mathbf{B}.\partial_t\mathbf{J}).\nonumber
\end{flalign}
This is the conservation law of optical chirality where $\mathbf{B}.\partial_t\mathbf{D}-\mathbf{D}.\partial_t\mathbf{B}$ is the \textit{optical chirality density} and $\mathbf{E}\times\partial_t\mathbf{D}+\mathbf{H}\times\partial_t\mathbf{B}$ is the \textit{optical chirality flow}\cite{lashkari2014energy}.

\section*{$5$D ELECTROMAGNETIC}
To generalize $4$-Maxwell's equations, \eqref{maxwell1} and \eqref{maxwell2}, to $5$-Minkowski space-time we use the same method as we did for $4$-EM field. First we write $F^{\alpha\beta}$ and $F_{\alpha\beta}$ matrices to define fields' vectors. For the new dimension we need two new vectors, $\mathbf{V}$ and $\mathbf{W}$, and two new scalars, $P$ and $R$, as new elements of these matrices. Notice these new vectors, like $\mathbf{E}$ and $\mathbf{D}$, and two new scalars are vectors and scalars in three dimentional spatial sub-space, not in $5$ dimentional space.
\begin{align}
\big(F^{\alpha\beta}\big):=
\begin{pmatrix}
 0   &  D_x &  D_y &  D_z &  R   \\
-D_x & 0    &  H_z &- H_y &  W_x \\
-D_y &- H_z &  0   &  H_x &  W_y \\
-D_z &  H_y &- H_x &  0   &  W_z \\
-R   &- W_x &- W_y &- W_z & 0
\end{pmatrix}
\end{align}
and
\begin{align}
\big(F_{\alpha\beta}\big):=\mu
\begin{pmatrix}
 0   & -E_x & -E_y & -E_z & -P   \\
 E_x &  0   &  B_z &- B_y &  V_x \\
 E_y &- B_z &  0   &  B_x &  V_y \\
 E_z &  B_y &- B_x &  0   &  V_z \\
 P   &- V_x &- V_y &- V_z & 0
\end{pmatrix}.
\end{align}
Then by defining a metric similar to \eqref{metric}
\begin{flalign}\label{metric5}
(g_{\alpha\beta}):=\text{diag}(-\frac{1}{\epsilon},\mu,\mu,\mu,\kappa)
\end{flalign}
we find new relations of $\mathbf{V}=\kappa \mathbf{W}$ and $P=R\kappa/(\epsilon\mu)$, additional to the pervious relations in \eqref{FRelations}. $\kappa>0$ is a new constant for the new spacelike dimension, $u:=x^4$. To be able to compare $\kappa$ with $\mu$, let the unit of measurement of $u$ be the same as spatial coordinates.\\
By applying Maxwell's equations, \eqref{maxwellF} and \eqref{maxwellFJ}, to $5$-Minkowski space-time we find $5$-Maxwell's equations
\begin{flalign}\label{5maxwell1}
\nabla.\mathbf{B}&=0\nonumber\\
\partial_t\mathbf{B}+\nabla\times \mathbf{E}&=0\nonumber\\
-\partial_u\mathbf{E}+\partial_t{\mathbf{V}}+\nabla P&=0\nonumber\\
\partial_u\mathbf{B}+\nabla\times\mathbf{V}&=0.
\end{flalign}
and
\begin{flalign}\label{5maxwell2}
\nabla.\mathbf{D}+\partial_u R&=4\pi J^0\nonumber\\
-\partial_t\mathbf{D}+\nabla\times \mathbf{H}+\partial_u\mathbf{W}&=4\pi\mathbf{J}\nonumber\\
\partial_tR+\nabla.\mathbf{W}&=-4\pi J^4.
\end{flalign}
Before going further we need to have an explanation to why a HELD has remained undetected which narrows down to explaining why the new terms in \eqref{5maxwell1} and \eqref{5maxwell2} have not produced detectable measures in the experiments. Most of such experiments have assured us that for example the vacuum is following $4$-Maxwell's equations, which means that the original $4$-Maxwell's equations have remained unaffected by these new terms of $5$-Maxwell's equations. We argue that a large enough $\kappa$ can omit the effects of the new HELD, $u$, from $5$-Maxwell's equations. This is because it can be shown that $\kappa$ is related to the group velocity of light in that dimension, which is so small compared to group velocity of light on the ordinary spatial directions to hinder any possible detectable effects. Given that light must carry information in a specific dimension for it to be detectable, small group velocity means inability to carry information in a logical time scale. To calculate the group velocity of light in the new direction one does not need to solve \eqref{5maxwell1} and \eqref{5maxwell2} for a plane wave, and only the metric itself can be used for this end. By using metric\eqref{metric5} the invariance quantity of space-time is
\begin{align}\label{sinvar}
ds^2=-\frac{1}{\epsilon}dt^2+\mu(dx^2+dy^2+dz^2)+\kappa du^2
\end{align}
where $ds/\sqrt{\mu}$ is the \textit{propper distance}. Light's path is on the null geodesics, $ds=0$, which means the group velocity of light on ordinary spatial dimensions is $dx/dt=1/\sqrt{\mu\epsilon}$ and along HELD is $du/dt=1/\sqrt{\kappa\epsilon}$. As long as group velocity of light on HELD is much smaller than that in the ordinary spatial dimensions, $du/dt\ll dx/dt$, then $\kappa$ is a large number compared to $\mu$, $\kappa\gg\mu$. This implies that our $4$-Minkowski space-time of vacuum is on the boundary of a $5$ dimensional counterpart, but as long as photons do not travel along the extra dimension, or they do it very slowly, we cannot observe the effects. We call this condition, $du/dt\ll dx/dt$, \textit{group velocity of light} or \textit{GVL} condition.\\
This concludes that one should search for possible effects of a HELD where $\kappa$ is smaller and the group velocity of light is in detectable zone. However, it contradicts our presumption that metric must be constant, because $\kappa$ is a component of the metric and we need to decrease its value.    So for the sake of consistency we reduce our assumption to the following: ``the metric of materials for monochromatic waves is a constant'', then for different regions with different materials the metric can be different, but inside those regions its components are some constans. This assumption is implicit in how we have used $\mu$ and $\epsilon$ for dispersive media. We call this \textit{partially constant metric} or \textit{PCM} assumption.\\
Notice that allowing the metric to change is not an optional feature but it is a requirement. This implies for a consistent theory we need to allow the metric to change everywhere. This is because if the metric was really partially constant then we could always transform the coordinates to keep $\kappa$ equal to $\mu$, which would make all the differences of the HELD with ordinary spatial dimensions trivial. Here we are interested in a local view of the HELD and we claim the group velocity of light in that dimension is too small to carry any information regardless of the unit we may choose. \\

\subsection*{AN EXAMPLE OF A POSSIBLE GLOBAL METRIC}
Let us make an example or an ``existence proof'' for clarfying how the metric must change where GVL condition can be satisfied. A \textit{Schwarzschild-like} metric in five dimensions is the simplest case\cite{petersen1999introduction}
\begin{flalign}\label{schw}
ds^2=-\Big(1-\frac{M^2}{r^2}\Big)dt^2+{\Big(1-\frac{M^2}{r^2}\Big)}^{-1}dr^2+r^2d\Omega^2_3.
\end{flalign}
Where $r$ is the HELD, $d\Omega^2_3$ is the metric on the unit sphere $S^3$ in the transverse space which locally would be our ordinary 3 dimenstional space, and $M$ is the radius of horizon. First of all, we need to find the locale time that a falling particle measures in this metric\cite{townsend1997black}. To do so we need to find its geodesic by solving $\nabla_pp=0$, where $p$ is the tangent vector to the trajectory of the falling particle, $p=p^\alpha\partial_\alpha=m dx^\alpha/d\tau\partial_\alpha$, $\tau$ is the locale time that it measures, and $m$ is its mass. But instead of solving geodesics equations, $\nabla_pp=0$, we can use a simpler method by looking at constants of motion. As long as initial tangent vector of the trajectory in $r-t$ plane has two components, we should have two constants for this motion, which are
\begin{flalign}\label{constEqu}
Q_1=\langle p,p\rangle\quad\textit{and} \quad Q_2=\langle k,p\rangle
\end{flalign}
where $k$ is a killing vector that satisfies
\begin{flalign}
\nabla_\mu k_\nu+\nabla_\nu k_\mu=0\quad\textit{or} \quad \langle v,\nabla_v k\rangle=0
\end{flalign}
for any vector $v$ and $\nabla$ is the covariant derivative. With these equations we can show
\begin{flalign}
\frac{dQ_1}{d\tau}=0\quad\textit{and}\quad\frac{dQ_2}{d\tau}=0.
\end{flalign}
The first constant is the minus square mass of the particle, $Q_1=-m^2$. Also the obvious killing vector of the metric \eqref{schw} is $k=d/dt$ then its constant would be energy, $Q_2=p_t=-m\varepsilon$. The first equation of \eqref{constEqu} is ${(p^r)}^2-{(p_t)}^2=-m^2(1-\frac{M^2}{r^2})$ and the second equation is $\varepsilon=(1-M^2/r^2)dt/d\tau$ which allow us to write the metric \eqref{schw} with locale time of the particle
\begin{flalign}
ds^2={\Big(1-\frac{M^2}{r^2}\Big)}^{-1}\left(-\varepsilon^2d\tau^2+dr^2\right)+r^2d\Omega^2_3.
\end{flalign}
Now, we are in a position that we can check the GVL condition. Locally we have $r^2d\Omega^2_3=dx^2+dy^2+dz^2$, $r>M$, then also $dr/d\tau=\varepsilon$ and $dx/d\tau=\varepsilon r{\left(r^2-M^2\right)}^{-1/2}$ for light, $ds=0$. This implies the condition $dr/d\tau<dx/d\tau$ will be satisfied. This example may be enough as a proof of existence of a global metric where GVL condition is satisfied, but there is a more interesting situation. Inside the horizon of this black hole, \eqref{schw}, $r$ is the timelike coordinate and $t$ is the spacelike one, hence the extra dimension in this scenario. By using the first equation of \eqref{constEqu}, which is $({dr/d\tau})^2=\varepsilon^2-1+\frac{M^2}{r^2}$, we can write \eqref{schw} as
\begin{flalign}\label{schwInside}
ds^2= -\frac{\varepsilon^2-1+\frac{M^2}{r^2}}{\Big(\frac{M^2}{r^2}-1\Big)}d\tau^2+\Big(\frac{M^2}{r^2}-1\Big)dt^2+r^2d\Omega^2_3.
\end{flalign}
Thus the GVL condition, $dt/d\tau<dx/d\tau$, inside the black hole for $r<M/\sqrt{2}$ will be satisfied. This situation is more interesting because we can consider crossing of particles through the horizon as the \textit{Big Bang}. This is because $dx/d\tau$ of light becomes infinity in that moment, at least in theory. Also a model like this directly predicts that falling particle inside this black hole can measure the redshift of other falling particles (see Supplementary Note), or in other worlds, these particles would see thier universe expanding! Additional to these facts, the group velocity of light on ordinary spatial dimensions, $dx/d\tau$, must be decelerating, which allows the \textit{Cosmological Principle} to be satisfied. A similar idea can be found here\cite{albrecht1999time}. \\\\
Comming back to our local view of extra dimension, the PCM assumption, we will need conversation law of optical chirality in $5$-Minkowski space-time. By applying $5$-Maxwell's equations \eqref{5maxwell1} and \eqref{5maxwell2}
\begin{flalign}
&\partial_t\big(\mathbf{B}.\partial_t\mathbf{D}-\mathbf{D}.\partial_t\mathbf{B}\big)\nonumber\\
&+\nabla.\big(\mathbf{B}\times\partial_t\mathbf{H}+\mathbf{E}\times\partial_t\mathbf{D}\big)\nonumber\\
&+\partial_u\big(\mathbf{W}.\partial_t\mathbf{B}\big)\refstepcounter{equation}\subeqn\label{vecchirlity1}\\
&=4\pi (\mathbf{J}.\partial_t\mathbf{B}-\mathbf{B}.\partial_t\mathbf{J})\nonumber\\
&\quad+\mathbf{W}.\partial_t\partial_u\mathbf{B}+\mathbf{B}.\partial_t\partial_u \mathbf{W}\subeqn\label{vecchirlity2}
\end{flalign}
where $\mathbf{W}.\partial_t\partial_u\mathbf{B}+\mathbf{B}.\partial_t\partial_u \mathbf{W}$ is a new term in $n=5$ for generation and anihilation of optical chirality inside a region, even in absence of a current-density vector, $\mathbf{J}$. This generalization for $n>4$ was made in a previous work\cite{lashkari2014energy}. This generalization equips us with a set of equations that predict specific physical measurables in $5$-EM that lack in $4$-EM. The next section is devoted to studying a potential test bed for detecting this HELD.

\section*{SUPERCONDUCTORS}
Despite our derivation of a new set of equations for $5$-EM, a place to apply these equations needs to be found. Here we suggest that type I superconductors can be a test bed for our hypothesis. This is because, as we will show, some formal equations that govern the physics of superconductors are the sub-equations of $5$-Maxwell's equations, \eqref{5maxwell1} and \eqref{5maxwell2}. This means that we can guess answers for the HELD, $u$, and by applying $5$-Maxwell's equations on them, we will derive the formal equations of superconductors. We start by London equations\cite{london1935electromagnetic,cyrot1992introduction,spaldin2010magnetic}
\begin{flalign}\label{london}
\frac{\partial \mathbf{J}_s}{\partial t}=\frac{n_s e^2}{m}\mathbf{E},\qquad
\nabla\times \mathbf{J}_s = -\frac{n_s e^2}{m}\mathbf{B}
\end{flalign}
where $\mathbf{J}_s$ is the superconducting current-density, $e$ is the charge of electron, $m$ is the electron's mass and $n_s$ is a constant. Interestingly we observe that if we make the following substitutions for solution of dimension $u$:  $P\rightarrow0,\quad\mathbf{J}\rightarrow0$ and
\begin{flalign}\label{londonpar}
&\mathbf{V}\rightarrow \frac{4\pi \kappa}{k_u}\mathbf{J}_s(t,x,y,z) e^{-ik_uu},
\ \mathbf{E}\rightarrow i\mathbf{E}(t,x,y,z) e^{-ik_uu}\nonumber\\
&\mathbf{B}\rightarrow i\mathbf{B}(t,x,y,z) e^{-ik_uu},\qquad
\frac{{k_u}^2}{4\pi\kappa}\rightarrow\frac{n_s e^2}{m}
\end{flalign}
into \eqref{5maxwell1} and \eqref{5maxwell2}, we retrieve London equations\eqref{london}. In other words by proper choices one can fit the London equations into $5$-Maxwell's equations. Since London equations only describe the Meissner effect, we need more evidence. The next set of equations that we investigate are Josephson's equations which were originally derived by \textit{Quantum Mechanical} considerations\cite{josephson1962possible,anderson1970josephson}.
\begin{flalign}\label{josephson}
J=J_c\sin \phi,\qquad
\frac{d \phi}{d t}=\frac{2ed}{\hbar}E_0
\end{flalign}
where the junction is in the $x$ direction, $d$ is a distance in that direction, $\hbar$ is the reduced Planck constant, $J$ is the current-density of the Josephson effect and $E_0d$ is the voltage across the Josephson junction. So again we propose an answer for dimension $u$ and substitute it in \eqref{5maxwell1} and \eqref{5maxwell2}
\begin{flalign}
&P\rightarrow0,\qquad \mathbf{B}\rightarrow0\nonumber\\
&E_y\rightarrow0,\qquad E_z\rightarrow0\nonumber\\
&V_y\rightarrow0,\qquad V_z\rightarrow0\nonumber\\
&J_y\rightarrow0,\qquad J_z\rightarrow0\nonumber
\end{flalign}
\begin{flalign}\label{jospar}
&V_x\rightarrow V_0 \sin \phi(t) e^{k^{'}_uu},\qquad
E_x\rightarrow E_0 \cos \phi(t) e^{k^{'}_uu}\nonumber\\
&J_x\rightarrow J e^{k^{'}_uu}
\end{flalign}
\begin{flalign}
\frac{k^{'}_u}{V_0}\rightarrow\frac{2ed}{\hbar},\qquad
\frac{1}{4\pi} \left(\epsilon {E_0}\omega_0+\frac{V_0 k^{'}_u}{\kappa}\right)\rightarrow J_c\label{josephpar}
\end{flalign}
then we can recover Josephson equations,\eqref{josephson}. Here $\omega_0=k^{'}_vE_0/V_0$ is the angular frequency of Josephson effect. To understanding the fact that electromagnetic field grows exponentially, $e^{k^{'}_uu}$, we need to remember that the metric is not constant generally, so we can argue this answer is just for a narrow vicinity of $u=0$, where the electromagnetic field changes as $e^{k^{'}_uu}$. We cannot, however, extend our assumption further, as we are not certain how the metric behaves in those locations. Additionally one can add $\mathbf{B}$ to equations, \eqref{josephson}, and show there are still sub-equations of \eqref{5maxwell1} and \eqref{5maxwell2}. This implies $5$-Maxwell's equations are capable of reproducing superconductor's
equations. This statement does not hold for $4$-Maxwell's equations. Despite all these, we still need a measurable effect, which is predicted by $5$-EM but is absent in $4$-Maxwell's equations. In the previous section we proposed that the conservation law of optical chirality is a good candidate for this purpose. In absence of current density it is predicted to be conserved by $4$-EM. However according to $5$-EM it may not be conserved even in lack of any current density.   

\section*{PROPOSED EXPERIMENT}
The groundwork for the proposed experiment is laid by considering two separate findings that the conservation of optical chirality behaves differently for $n=4$ than $n=5$ and the conjecture that superconductors are possible candidates for housing a detectable HELD. In this section we try to combine them to propose an experiment inside a Josephson junction to measure optical chirality's effect. Assume an incident plane wave inside the transparent dielectric barrier of a Josephson junction
\begin{flalign}
&\mathbf{B}_i=\mathbf{B}_0e^{i(\omega t-\mathbf{k}.\mathbf{x})},\qquad
\mathbf{E}_i=\mathbf{E}_0e^{i(\omega t-\mathbf{k}.\mathbf{x})}\nonumber\\
&\mathbf{V}_i=\mathbf{J}_i=0,\qquad\qquad P=0.
\end{flalign}
We rename the parameters of \eqref{jospar} to $\mathbf{V}_j$, $\mathbf{B}_j$, $\mathbf{E}_j$ and $\mathbf{J}_j$. So total electromagnetic field inside barrier would be
$
\mathbf{V}=\mathbf{V}_i+\mathbf{V}_j,\
\mathbf{B}=\mathbf{B}_i+\mathbf{B}_j,\
\mathbf{E}=\mathbf{E}_i+\mathbf{E}_j\
\text{and}\
\mathbf{J}=\mathbf{J}_i+\mathbf{J}_j.
$
Here we need to make an argument that no optical chirality exits the boundaries of $t$ and $u$, which means no optical chirality accumulates inside the barrier, so it does not go out of the boundaries of t. Also no optical chirality propagates along $u$. The GVL condition predicts no information gets carried along the extra dimension. On the other hand, optical chirality as the carrier of handedness or helicity of light carries information. As a result one does not expect the optical chirality to propagate along $u$ to avoid any information carriage. As was argued before we expect $\kappa$ to decrease inside superconductors but it still is large enough for GVL to be a valid condition. This argument implies that after integrating \eqref{vecchirlity1} over the region of barrier, the derivatives of $t$ and $u$ become zero. After this step the optical chirality inside the Josephson junction is measured by \eqref{vecchirlity2}
\begin{flalign}\setcounter{subeqn}{0}
c&=\frac{\mu}{v^2}\int\int\oint \Big(\mathbf{B}\times\partial_t\mathbf{H}+\mathbf{E}\times\partial_t\mathbf{D} \Big).d\mathbf{S}  dtdu\refstepcounter{equation}\subeqn\label{measurable}\\
&=\frac{\mu}{v^2}\int\int\int  \Big(4\pi (\mathbf{J}.\partial_t\mathbf{B}-\mathbf{B}.\partial_t\mathbf{J})\nonumber\\
&\qquad\qquad+\mathbf{W}.\partial_t\partial_u\mathbf{B}
+\mathbf{B}.\partial_t\partial_u \mathbf{W}\Big) d\mathcal{V}dtdu.\subeqn\label{calculatable}
\end{flalign}
where $c$ is the optical chirality inside the barrier. Cancelling the constants, the first line is a measurable quantity, \eqref{measurable}, and the second one, \eqref{calculatable}, depends on which theory we use, $n=4$ or $n=5$. If we calculate the integral in \eqref{calculatable} we will get:
\begin{flalign}
\int_0^{2\pi/\omega}\oint &\Big(\mathbf{B}\times\partial_t\mathbf{H}+\mathbf{E}\times\partial_t\mathbf{D} \Big).d\mathbf{S}  dt\nonumber\\
=&\mathbf{B}_0.\hat{\mathbf{i}}\frac{4\pi a}{k_y}\left(e^{-ik_yl_y}-1\right)\nonumber\\
&\times\Bigg(\frac{ i(\omega^2J_c+{\omega_0}^2J_d)\sin \left(2\pi\frac{\omega_0}{\omega}\right)}{\omega^2-{\omega_0}^2}\nonumber\\
&\quad+\frac{\omega\omega_0(J_c+J_d)\big(1-\cos \left(2\pi\frac{\omega_0}{\omega}\right)\big)}{\omega^2-{\omega_0}^2}\Bigg)
\end{flalign}
where $k_x$ and $k_z$ are zero, $a$ is the incident area of the light that is going through the barrier, $l_y$ is the distance that light is inside the barrier and $J_d$ is a constant. $\hat{\mathbf{i}}$ is the Cartesian unit vector in $x$ direction and appears because the junction is in this direction. for the case of $4$-EM one can easily calculate that
\begin{flalign}
J_d\rightarrow  J_c
\end{flalign}
and in the case of $n=5$ one obtains:
\begin{flalign}
J_d\rightarrow J_c-\frac{k^{'}_u V_0}{4\pi\kappa}.
\end{flalign}
Clearly by measuring the generation and annihilation of optical chirality of a plane wave inside a barrier of a Josephson junction, one can rule out one in favor of the other.

\section*{CONCLUSION}
In summary, we extended Maxwell's equations to $5$-Minkowski space-time, and showed that London and Josephson equations of superconductors can be ``fitted'' into them. Following this method we proposed a new experiment to test the existence of a new lengthlike dimension. We showed that conservation or non-conservation of optical chirality inside a Josephson barrier under specific experimental conditions may act as a measure for the existence of an extra lengthlike dimension. Also we described under what conditions such an additional dimension avoids detection in the vacuum.

\bibliography{superconductivity}
\bibliographystyle{unsrt}

\section*{Supplementary Note: Redshift Inside The Black Hole}
Here we show that a particle falling inside a Schwarzschild-like balck hole, \eqref{schwInside}, would see redshift of other particles which are falling too. In order to find the light redshift, which is exchanged between two falling particles, we need to solve \eqref{constEqu} equations for a photon. The first one is
\begin{flalign}\label{lightTraj}
0= -\frac{\varepsilon^2-1+\frac{M^2}{r^2}}{\Big(\frac{M^2}{r^2}-1\Big)}{\left(\frac{d\tau}{d\lambda}\right)}^2+\Big(\frac{M^2}{r^2}-1\Big){\left(\frac{dt}{d\lambda}\right)}^2+{\left(\frac{dx}{d\lambda}\right)}^2
\end{flalign}
where $\lambda$ is the parameter of light's trajectory and we kept $dx$ non-zero to let the light move in ordinary spatial space. The second equation is the same as the particle's, $\varepsilon^\prime=(1-M^2/r^2)dt/d\lambda$, where $\varepsilon^\prime$ is the energy of the photon. After we kept $dx$ non-zero, we need another constant to solve \eqref{lightTraj}, so we can use $d/dx$ killing vector to find an additional equation, $p_x=g_{xx}dx/d\lambda=dx/d\lambda$, where $p_x$ is the momentum of the photon on $x$ direction. By using all these three equations we can find
\begin{flalign}\label{redshifInR}
{\left(\frac{d\tau}{dx}\right)}^2=\frac{M^2+\left({\left(\frac{\varepsilon^\prime}{p_x}\right)}^2-1\right)r^2}{M^2+(\varepsilon^2-1)r^2}.
\end{flalign}
$\varepsilon^\prime/p_x$ is a constant that explains the slope of light trajectory and we name it $\alpha=\varepsilon^\prime/p_x$.  Also, we can find the relation between $r$ and $x$ by integrating ${(dr/dx)}^2=(M^2+(\alpha^2-1)r^2)/r^2$, where this equation is found by the first equation of \eqref{constEqu} and \eqref{redshifInR}. With $dr/dx<0$ condition, because photons are falling into the black hole too, the solution is $M^2-(1-\alpha^2)r^2={(1-\alpha^2)}^2{(X+x)}^2$ where $X>0$ is a constant of integration. Here we choose $\alpha^2\ll 1$ or ${(\varepsilon^\prime)}^2\ll {(p_x)}^2$ because the GVL condition forces most information to propagate by light in $x$ direction instead of the extra dimension, $t$. By replacing this equation into \eqref{redshifInR} we have
\begin{flalign}
\frac{d\tau}{dx}=\frac{(1-\alpha^2)(X+x)}{\sqrt{M^2\left(\frac{\varepsilon^2-\alpha^2}{1-\alpha^2}\right)-(\varepsilon^2-1)(1-\alpha^2){(X+x)}^2}}.
\end{flalign}
Now we can calculate the redshift, which is defined by
\begin{flalign}\label{redshift}
z=\frac{d\tau}{d\tau_0}-1=\frac{d\tau/dx}{d\tau_0/dx}-1
\end{flalign}
where $d\tau_0/dx=d\tau/dx|_{x=0}$ is a constant. \eqref{redshift} is an increasing function of $x$ if $\varepsilon^2>\alpha^2$. This implies that for photons whose $\alpha$ is smaller than $\epsilon$ of our falling universe, which as we explained include most of photons because of the GVL condition, we will see this redshift.

\end{document}